\documentclass{IEEEtran}
\usepackage{proof}
\usepackage[utf8]{inputenc}
\usepackage[spanish,english]{babel}
\usepackage[T1]{fontenc}
\usepackage{amsmath}
\usepackage{amssymb}
\usepackage[]{graphicx}
\usepackage{paralist}
\usepackage{enumitem}
\usepackage{color}
\usepackage{xspace}
\usepackage{url}
\usepackage[numbers]{natbib}
\usepackage{csquotes}
\usepackage{mathrsfs}

\usepackage{multirow}
\usepackage[table,xcdraw]{xcolor}

\newcommand{\HTTP}{\textsc{http}\xspace}
\newcommand{\HTML}{\textsc{html}\xspace}
\newcommand{\WAF}{\textsc{waf}\xspace}
\newcommand{\OWASP}{\textsc{owasp}\xspace}
\newcommand{\CSIC}{\textsc{csic2010}\xspace}
\newcommand{\CRS}{\textsc{owasp crs}\xspace}
\newcommand{\DARPA}{\textsc{darpa 1999}\xspace}
\newcommand{\SQL}{\textsc{sql}\xspace}

\newcommand{\URI}{\textsc{uri}\xspace}
\newcommand{\URL}{\textsc{url}\xspace}

\newcommand{\TCP}{\textsc{tcp}\xspace}
\newcommand{\XSS}{\textsc{xss}\xspace}
\newcommand{\SQLi}{\textsc{SQLInjection}\xspace}
\newcommand{\DBF}{\textsc{dbf}\xspace}
\newcommand{\NLP}{\textsc{nlp}\xspace}
\newcommand{\IP}{\textsc{ip}\xspace}
\newcommand{\PKDD}{\textsc{ecml/pkdd2007}\xspace}
\newcommand{\FING}{\textsc{drupal}\xspace}
\newcommand{\IDS}{\textsc{ids}\xspace}
\newcommand{\CRLF}{\textsc{crlf}\xspace}
\newcommand{\GET}{\textsc{get}\xspace}
\newcommand{\POST}{\textsc{post}\xspace}
\newcommand{\ModSecurity}{\textsc{ModSecurity}\xspace}

\newcommand{\R}{\mathbb{R}}
\newcommand{\A}{\mathscr{A}}

\newcommand{\C}{\mathscr{C}}

\newcommand{\T}{\mathscr{T}}

\newcommand{\N}{\mathbb{N}}
\newcommand{\Distrib}{\mathscr{D}}
\newcommand{\dist}{\textit{dist}}

\newcommand{\cluster}{\textit{C}}
\newcommand{\meanDist}{\overline{dist}}
\newcommand{\stdDist}{\textit{std}}

\newcommand{\osst}{\osst}

\newlength{\bcextramargin}
\newenvironment{changemargin}[2]{\begin{list}{}{%
\setlength{\topsep}{0pt}%
\setlength{\leftmargin}{0pt}%
\setlength{\rightmargin}{0pt}%
\setlength{\listparindent}{\parindent}%
\setlength{\itemindent}{\parindent}%
\setlength{\parsep}{0pt plus 1pt}%
\addtolength{\leftmargin}{#1}%
\addtolength{\rightmargin}{#2}%
}\item }{\end{list}} 

\newcommand{\actdefsection}[1]{
\begin{changemargin}{-\bcextramargin}{0pt}
\vspace{1ex}
\noindent
\textbf{{#1}}
\end{changemargin}
}

\newif\iffull\fullfalse

\begin{document}
\title{Machine learning-assisted virtual patching of \\ web applications}

%
\author{
\IEEEauthorblockN{
    Gustavo Betarte\IEEEauthorrefmark{1}\IEEEauthorrefmark{2},
    Eduardo Gim\'enez\IEEEauthorrefmark{2},
    Rodrigo Mart\'inez\IEEEauthorrefmark{1}\IEEEauthorrefmark{2} and
    \'Alvaro Pardo\IEEEauthorrefmark{3} }    \\
  \IEEEauthorblockA{
    \IEEEauthorrefmark{1}Instituto de Computaci\'on, Facultad de Ingenier\'ia
      Universidad de la Rep\'ublica, Uruguay \\
      Email: [gustun,rodmart]@fing.edu.uy}  \\
   \IEEEauthorblockA{
    \IEEEauthorrefmark{2}Tilsor SA, Uruguay \\
      Email: [gbetarte,egimenez,rmartinez]@tilsor.com.uy}  \\
   \IEEEauthorblockA{
    \IEEEauthorrefmark{3}Departamento de Ingenier\'ia El\'ectrica,
      Facultad de Ingenier\'ia y Tecnolog\'ias \\ Universidad Cat\'olica del Uruguay, Uruguay \\
      Email: apardo@ucu.edu.uy}
   \thanks{This research was partially funded by ICT4V (https://www.ict4v.org).
The results reported here are the result of work carried out in the context of WAFINTL, a project focused on WAF technology. }   
}

\maketitle

\begin{abstract}


Web applications are permanently being exposed to attacks that exploit their vulnerabilities. In this work we investigate the application of machine learning techniques to leverage Web Application Firewall (\WAF), a technology that is used to detect and prevent attacks. We propose a combined approach of machine learning models, based on one-class classification and n-gram analysis, to enhance the detection and accuracy capabilities of \ModSecurity, an open source and widely used \WAF.

The results are promising and outperform \ModSecurity when configured with the \OWASP Core Rule Set, the baseline configuration setting of a widely deployed, rule-based \WAF technology. The proposed solution, combining both approaches, allow us to deploy a \WAF when no training data for the application is available (using one-class classification), and an improved one using n-grams when training data is available.   

\end{abstract}


%

\section{Introduction}
A web application is a piece of software, based on a client-server architecture, that embodies a coordinated set of functions. The information flowing between the client, which runs on the user's web browser, and the application server is transmitted using the \HTTP protocol.

By its very nature web applications are designed to be exposed, therefore they are available to any individual, or artifact, with capabilities to access the Internet. Thus, web applications are a primary target for any attacker who wants to unauthorizedly access the information they handle or to place baits (e.g., fake URLs to his own site) to lure honest users. It is quite usual for the code of a web application to contain vulnerabilities like the ones listed and described in the \textsc{owasp top 10}~\cite{owaspTop10}. The work reported in this paper focuses on preventing the abuse of web applications. 

A Web Application Firewall (\WAF) is a piece of software that intercepts and inspects all the traffic between the web server and its clients, searching for attacks inside the \HTTP packet contents. Once recognized, the suspicious packets may be processed in a different, secure way, for instance being logged, suppressed or derived to a honeypot application. 

\ModSecurity~\cite{ModSecurity} is an open source, widely used \WAF enabling real-time web application monitoring, logging and access control. The actions \ModSecurity undertakes are driven by rules that specify, by means of regular expressions, the contents of the \HTTP packets to be spotted. 
\ModSecurity offers a default set of rules, known as the \OWASP Core Rule Set (\CRS)~\cite{owasp-crs}, for tackling the most usual vulnerabilities included in \cite{owaspTop10}. However, an approach only based on rules also has some drawbacks: rules are static and rigid by nature, so the \CRS usually produces a rather high rate of false positives, which in some cases may be close to 40\%~\cite{ModSecurityFalseRate}. Rule tuning is a time consuming and error prone task that has to be manually performed for each specific web application. In traditional networks firewalls and \IDS, the approach based on rules has been successfully complemented with other machine learning-based tools, anomaly detection and other statistical approaches which provide higher levels of flexibility and adaptability. Those approaches take advantage of sample data, from which the normal behavior of the web application can be learned, in order to spot suspicious situations which fall out of this nominal use (anomalies), and which could correspond to on-going attacks. Our final objective is to improve \ModSecurity with such anomaly detection techniques.

The design of a statistically-based \WAF may be built applying the knowledge that a security expert has on both the application itself and the current state of the art on attack techniques. Alternatively, a description of the expected normal behavior of the web application may be built from legitimate users' input. In the first approach the \WAF is trained to recognize payloads that are instances of known attacks and any input which cannot be recognized as an attack is deemed to be a valid input. The second approach adopts the symmetric point of view: the \WAF learns the expected inputs of the web application and any other bias from the expected values is deemed as a potential attack. As we shall show in what follows both approaches can be composed to recognize a broader class of web application attacks.

\paragraph*{Organization of the paper}
The structure of the rest of the paper is as follows: Section~\ref{section:background} provides some background and describes the approach we have taken to enhance \ModSecurity virtual patching capabilities with machine-learning techniques. Section \ref{section:learnmodels} presents the two complementary learning models that we use to ground a statistical \WAF. Then, in Section~\ref{section:experiments} we describe and discuss the outcomes of the experiments we have performed. Section~\ref{section:relwork} reviews related work. Further work and conclusions are presented in Section~\ref{section:conclusion}.


\section{Background and problem statement}
\label{section:background}

\subsection{Web application security and virtual patching}
Usually, successful attacks result from exploiting vulnerabilities present in the web application. The inspection and fixing of the application's code, however, might not always be feasible. Critical applications, for instance, can not be off-line until the bug has been fixed. Moreover, it might well be the case that the source code of the application is neither accessible nor longer available or patchable. We are interested in the use of security \textit{virtual patching} as a remediation technique. In particular, we seek to enhance the functioning of the widely deployed \WAF called \ModSecurity. This virtual patcher can be used to protect any web application and its working relies on standardized and public protocols such as \HTTP.  


Traditional network firewalls are designed to perform packet inspection at the network level. Web applications, however, exchange messages in higher layers, mostly \HTTP, that traditional firewalls are not designed to understand. \ModSecurity has been specifically designed to protect web applications with the ability to detect and identify patterns that could result in an attack against the application. It is able to intercept a broad class of attacks, including, among others, \XSS (cross-site scripting), any form of code injection and path traversal. \ModSecurity has two modes of operation: detection and prevention. In the first mode, records are generated for each detected potential attack and used to monitor specific rules. Normally, this mode is used in what is called learning phase.  The second mode is when the \WAF is really useful: by correctly configuring rules and directives it is able to block potential malicious Web traffic. The core of \ModSecurity implements a rule engine which can be set to execute in each transaction of the application.

The use of \ModSecurity provides a first line of defense against attacks and allows the detection of security problems using a rule-based system. In other words, it is capable of enforcing a negative security policy by blacklisting requests that are filtered by those rules. This frequently results in the generation of false alarms and requires a continuous update of those rules. Failure to do so may result in limited or no protection against new (zero-day) attacks and the generation of too many false positives,  which must be adjusted by adding exceptions to the rules for each application. These adjustments are strictly necessary in order to set to work the firewall, otherwise the number of false positives generated can be so high that basically results in a denial of service. However, the management of rule configurations for \ModSecurity is a complex and error-prone task. In particular, the writing and maintenance of these configurations requires mastering a low level language with an intricate syntax in which the modeling of complex attacks involves several rules possibly residing in multiple configuration files.

\subsection{Learning attack detection}
\ModSecurity is usually configured to work using a negative security policy because defining the rules that describe normal behavior of the application to protect  is an almost impossible task in real life. The problem with this operational mode is the high amount of false positives generated and the amount of work needed during the learning phase. The main objective of the research reported here is to enhance \ModSecurity operation by using machine learning techniques to adapt its behavior to that of the application that is protecting. Depending on the operational context of the target application one may consider alternative learning scenarios. 

The problem of protecting a web application can be addressed using two different automated learning approaches: as a \textit{classification} problem and as an \textit{anomaly detection} problem.  In the first approach protection of the application amounts to classify a new given input using a model that has been constructed from a previous training phase based on supervised learning and using a collection of inputs, each one labeled either as a valid input or as some kind of attack from an attack taxonomy. However, the ideal situation of having available a labeled dataset of application requests that represent valid and attack behavior of a specific application is not always possible.
The second approach focuses on characterizing which are the expected (valid) input values for the web application considering any anomalous input supplied to the application as a potential attack. Protecting the application therefore is equivalent to outlier detection from a training dataset of valid inputs.


The anomaly detection approach has several advantages. First, as any behavior different from the usual one is considered problematic, it can potentially make it possible to detect zero-days attacks. Additionally, training the \WAF only requires a collection of \HTTP requests produced by friendly users and no need for labelling the \HTTP requests. However, as noticed in \cite{closeLookToNgrams}, this approach cannot provide an answer to all possible attack scenarios. This is because some of the web application input data may be random data by its very nature. Consider for instance a web application parameter carrying a user password. It can be made of any combination of symbols, and in principle all combination should be equally likely. In such cases there is no hope to find a strong language signature for the field, nor to reject a chosen password on the basis that it is not frequently used. Furthermore, for security reasons, it would not be advisable for the \WAF to store information about the distribution of such sensible parameters. In those cases, where there is no a priori expected behavior, it is necessary to adopt a symmetric approach, and search for attack signatures in the field according to the current state of the art. This amounts to search for carefully selected tokens which have been extracted from a labeled training dataset and representing the knowledge of the security expert on the current attack vectors. This technique deems the \HTTP request as valid by default when it cannot be recognized as an attack according to the previous training.

\subsection{Improving \ModSecurity detection capabilities}
\label{improving_modsecurity}
We have been experimenting with different learning techniques to enhance the detection and accuracy capabilities of \ModSecurity. We have pursued a combined approach that we now proceed to explain.

First, we have experimented with a mechanism that integrates one-class classification and \ModSecurity configured using the \CRS rules out of the box. This basically combines two experts with the objective of classifying a request. When both experts agree (both say valid or both say attack) then the result is straightforward. 
In the case the one-class approach classifies a request as an attack, given that the \CRS rules have the know-how on attacks, we prioritize its answer (so if \ModSecurity classifies the request as valid then is valid). This integration decision also allows us to have a well known mechanism to train our \WAF in the case we found a false positive. Basically we need to tell \ModSecurity rules that this request is not an attack. This training mechanism is exactly the same as the one used with the \CRS nowadays. Finally, in the case that one-class classifies the request as valid and \ModSecurity as an attack we prioritize one-class since \CRS is known to have high false positive rates. 
This approach has shown to adapt quite well to the scenario where there is not available a specific training dataset for the application.  We train the one-class classifier using several datasets and the resulting classification model can be used to protect different web applications. The main advantage is that it is easy to deploy and capable of adapting to changes in the web application.


However, the detection capabilities provided by the one-class approach do not adapt so well to prevent both zero-day attacks and attacks that exploit specific vulnerabilities of a application, in particular those involving suspicious input. Thus, as a complementary solution, we have experimented with techniques that use high-order n-grams up to some $n$ to provide anomaly detection capabilities based on the expected language signature of the input fields of the application. The n-gram approach requires to have an application specific dataset with valid request to train its model. By modeling the language signature of each attribute of the web application it is possible to recognize known attacks with good precision and also to detect zero-days attacks.

In summary, if we need to fast deploy a \WAF to protect a web application without having a specific dataset or the web application changes constantly (e.g. public website powered by a content manager), we propose to use the one-class approach. We will call this Scenario I, when we want to protect a web application without having any specific training data. In this case we would like to address the following question: {\it Is it possible to build an attack detection system learning from training data collected from other web applications?}  A second question in this scenario it is related to \ModSecurity: {\it Can we improve the results of \ModSecurity using machine learning methods?} That is, can we reduce the number of false positives (FP) generated by \ModSecurity? In the scenario where we need to protect a business critical web application where high levels of security are required, and the application's changes are controlled, we can deploy the n-gram approach. In this second scenario, Scenario II, it shall be required to have a testing phase before each new deployment of the application, so specific training data will be available. This training data shall be used to train and update the n-gram model before the application's go-live. In this scenario, 
{\it we would like to understand the attainable performance of machine learning methods against that of \ModSecurity}.

\section{Learning Models}
\label{section:learnmodels}

We now turn to the description of the models proposed for enhancing the capabilities of \ModSecurity using machine learning techniques. These models follow a common pattern that we proceed to describe.

\subsubsection{Training set.} In web applications information between the client and the server is exchanged using the \HTTP protocol. This protocol is based in a request/response model using plain-text (ASCII) messages. The model is built up from a training set $\T$ of normal \HTTP requests that have been previously recorded. The set $\T$ is assumed to be representative of legitimate traffic. An \HTTP request is just a string with the following structure: a request header (specifying the method to be applied, the \URI on which it is applied and the protocol version), a collection of header fields of the form \textit{field=value} and possibly a request body. 

\subsubsection{Pre-processing} When non-ASCII information has to be exchanged between the client and the server the same encoding has to be employed. Different types of encoding might be used (e.g. URL encode \cite{berners1998rfc}). Before building the statistical model, the requests are first pre-processed to decode the information they contain. This might also involve parsing part of the \HTTP request structure or removing parts of it that are considered useless for the model. How granular is the parsing processing of the request structure constitutes a first choice in the model design.

\subsubsection{Feature extraction} Then, a collection of features are extracted from the request. These features are related to the occurrences of a collection $\A$ of substrings or \textsc{tokens} in the request. 
The criteria used to select these tokens is a second design choice of the model. We experimented with two different criteria: 
\begin{inparaenum}[i)]
\item using a fixed collection of words determined by a security expert from the current state of the art in web attacks, and 
\item automatically computing them from the training set (n-grams) 
\end{inparaenum}. This gives rise to an internal representation of the requests in terms of the numbers resulting from this counting. Such representation can be conceived as a vector of numbers, where each position of the vector contains the number associated to a given token, or more generally, as a mapping associating numbers to tokens (bag of words). 

\subsubsection{Model computation} The model is a probabilistic distribution for the elements of the internal representation estimated from those computed for the requests in the training set $\T$. The algorithm to be used for computing this distribution is a third choice to be made in the model design. This distribution provides the \textsl{signature} of the web application, which characterizes the normal \HTTP requests exchanged with the application. From a geometrical perspective, the model can also be conceived as a hyper-volume containing all the points (vectors) corresponding to the internal representation of a normal \HTTP request.

\subsubsection{Request classification} The \WAF uses a model $M$ to analyze incoming \HTTP requests on-line, one by one, as they are received, and classify them into two classes: normal and abnormal requests. Abnormal requests are supposed to be an attack, and hence processed differently according to the \WAF policy (logged, filtered, etc). In order to test a given \HTTP request $r$, the \WAF computes the internal representation of $r$. Then, a \textsl{score} $s_{r}=\dist(r,M)$ is computed for the request, using a distance function $\textit{dist}$ which measures how far the actual signature of $r$ is from the expected signature provided by the model $M$. A score of 0 means that the \HTTP request contents completely matches the expected signature. The higher the score, the less the field contents meets the expected frequency distribution. Another design choices are the distance and the criterion $\C(M,r)$ to be used for deciding whether the distance to the signature characterizing the application is far enough to deem the request $r$ as anomalous. Usually, this criterion relies on the distribution of distances that can be drawn from the training set $\T$, by computing the score of each individual training request.

The rest of this section is devoted to describe how the general pattern just described has been instantiated into two different models: one-class classification and n-gram analysis.


\subsection{One-Class Classification}
One-class classification \cite{khan2009survey} assumes that training information is only available for one of the classes; in our case the class of valid requests, this is also called \textit{novelty detection}\cite{pimentel2014review}. Hence, the problem of one-class classification is to learn the behavior of valid requests and identify attacks as deviations from the learned behavior for the valid class. That is, the goal is to define a boundary around the valid class maximizing the number of valid requests inside the boundary and minimizing the number of non valid instances inside it. Valid instances that fall outside of the defined boundary are false positives of the classifier.  

To build such a classifier we must define the features that are supposed to capture the properties of common known web application attacks (e.g. \SQLi, \XSS, among others). We shall rely on the the experience of a security expert  to do that. 
In~\ref{parag:features} we describe the features that will be used to characterize the request. 


\subsubsection{Pre-processing} 
In this phase, besides performing decodification, we also filter the request headers
that are used to exchange contextual information between the user-agent and the server. All information contained in headers that are specific to the protocol, such as cookies, proxies and \IP, which do not represent user behavior and should not be considered to infer application behavior, are filtered out.

\subsubsection{Feature Computation} \label{parag:features} Several known attacks make use of specially crafted inputs to force the server behave in a non expected manner. Most of these inputs use as part of the attack special characters, like for instance  $.$ $,$ $;$ $<$ $>$, or special substrings like for instance \textit{select}, \textit{alert}, \textit{passwd}. We propose to use these special characters and substrings to capture the characteristics of the request. This is basically a bag-of-words model where each document (in our case each request) is represented as a bag of its {\it words} (the special characters and substrings mentioned above). The bag-of-words feature vectors are calculated counting the appearance of each word in the request. In this way, each request $r$ is represented with a vector $x_r$ where each element in this vector counts the number of times each word is observed in $r$. To select the set of words, we studied valid and attack requests, and applied the knowledge of a security expert. In Table \ref{tab:selectedFeatures} we list all the words used. Since we are modeling the valid class, we expect that few of those words will be present in each request.  As we are focused on user input, we analyze the query string of the URL, the request body and the requests headers, all information that could be modified by the user of the application. 

\begin{table}[]
\centering
\caption{Selected special words by the security specialist}
\label{tab:selectedFeatures}
\begin{tabular}{lllll}
\textless             & ../     & alert           & exec        & password    \\
\textless\textgreater & '       & alter           & from        & path/child  \\
\textless!--          & “       & and             & href        & script      \\
=                     & (       & bash\_history   & \#include   & select      \\
\textgreater          & )       & between         & insert      & shell       \\
|                     & \$       & /c              & into        & table       \\
||                    & *       & cmd             & javascript: & union       \\
-                     & */      & cn=             & mail=       & upper       \\
--\textgreater        & \&      & commit          & objectclass & url=        \\
;                     & +       & count           & onmouseover & User-Agent: \\
:                     & \%00    & -craw           & or          & where       \\
/                     & \%0a    & document.cookie & order       & winnt       \\
/*                    & Accept: & etc/passwd      & passwd      &            
\end{tabular}
\end{table}

 
\subsubsection{Model computation}
The idea is to detect attacks as deviations from normality using a one-class classifier; a request will be classified as an attack if it is far from the observed valid class. A well known method to build a one-class classifier is to estimate the {\textit probability density function (pdf)} of the training data and to estimate a threshold to segment normal and abnormal instances. Among the existing methods to estimate the probability density function, Gaussian Mixture Models (GMM) is one of the most common one. If $x$ is a feature vector, it is assumed to be an observation of a random variable $X$ with a pdf: 
\begin{equation}
	p(x) = \sum_{k=1}^n P_i {\cal N}(x;\mu_k,\Sigma_k),
\end{equation}
 
\noindent where $n$ is the number of gaussians, $P_k$, $\mu_k$ and $\Sigma_k$ the weight, mean vector and covariance matrix of component $k$ respectively and ${\cal N}$ a Gaussian pdf. The Expectation Maximization (EM) algorithm \cite{em-dempster-77} can be used to estimate the parameters of the GMM (mean vector $\mu_k$, covariance matrices $\Sigma_k$ and weights $P_k$). In our case, each component of the obtained GMM constitutes a cluster that captures the distribution of valid requests. We also use EM to estimate $n$, the number of components (clusters). 


Once we have the GMM of valid class, in order to classify an instance into valid or attack, we need to compute the distance to each component of the GMM (cluster) and apply a threshold on this distance. 

In this work we use the Mahalanobis distance since it measures the distance of a feature vector to a distribution. In this case, the distribution corresponds to a component of the GMM ($\cluster_k$), so the distance is defined as:
\begin{equation}
	\dist(x_i,\cluster_k) = \sqrt{(x_i - \mu_k) \Sigma_k^{-1} (x_i - \mu_k)}
\end{equation}

\noindent If $x_i$ correspond to the mean of the distribution then the distance is 0, and increases when the point moves away taking into account the standard deviation of the distribution. If one of the dimensions has a standard deviation of $0$ in the distribution, the Mahalanobis distance could not be calculated. For this reason, we adjust the covariance matrix by adding a regularization term to it $\epsilon*\textit{Id}$, where $\epsilon$ is the smallest standard deviation in $diag(\Sigma_i)$ different from $0$ and $Id$ is the identity matrix. Intuitively this regularization allows the observation of new values on requests not seen in $\T$. 

\subsubsection{Request Classification} In this approach, the score $s_{r}$ is a vector where each position corresponds to the $\dist(r,\cluster_k)$. Before being able to do request classification, we need to estimate the threshold of each cluster. The threshold of the cluster is defined by analyzing the distribution of the distances of each instance of $\T$ that where assigned to the cluster. This transforms the multi-dimensional space of features into a real number corresponding to the distance of the request to the the cluster. Given that all requests assigned to a cluster (as a result of using EM) can be represented by a component of the GMM, we approximate the distribution of teh distances with a normal distribution. We define $\meanDist_k$ to denote the mean of all distance of $x_i$ to $\cluster_k$ where $x_i \in \cluster_k$ and $\stdDist_k$ to denote the standard deviation of the distances of $x_i \in \cluster_k$. Finally, the threshold $t_{\cluster_k}$ is defined as follows:
\begin{equation}
	t_{\cluster_k} = \lambda[  \meanDist_k + 10 * \stdDist_k ], \lambda \in (0,1] 
	\label{eq:one-class-threshold}
\end{equation}


The use of $\meanDist$ and $\stdDist$ makes it possible to have in the model a specific threshold for each cluster that depends directly on how the instances that belong to that cluster behave. The constant factor $10$ was derived using different dataset in a way that when $\lambda = 1$ the distance for all $x_i \in \cluster_k$ is less than $t_{\cluster_k}$. Varying the threshold, by multiplying it by a constant $\lambda$, let us change the model's precision. Even if there exists a specific threshold for each cluster, the $\lambda$ constant is the same for all clusters. This allows us to have only one parameter to control the precision of the whole model. As $\lambda$ grows more valid requests fall into the clusters, but also more attacks. When $\lambda$ decreases, the false positive rates increase, but more attacks are missed. This parameter allows us to have different operational points. 

Before deploying this approach to protect a web application, we need to define which operational point to use. The best scenario corresponds to have a labeled dataset with valid requests and attacks. In this scenario is possible to adjust this threshold to its best operational point. In real life applications, our best scenario is having examples of only normal behavior. In this case, one metric to define the operational point could be the amount of false positives that we are willing to accept. The worst scenario is when we do not have even an application specific dataset. For these scenario, we propose to use as training dataset a mixture of several datasets from different web applications. In Section \ref{section:experiments} we will present the results of this approach (Scenario I) to show that good performance scores can be achieved.


After training the model, the classification of a new request $r$ starts by calculating the score $s_{r}$. This results in a vector where each position corresponds to the Mahalanobis distance of $r$ to $\cluster_k$. The criterion $\C(M,r)$ for this approach corresponds to compare the score $s_{r}$ with the corresponding threshold for each cluster. If any of the scores is lower than the threshold the request is classified as normal, otherwise is classified as an attack.

\subsection{Anomaly detection using n-grams}
\label{subsection:anomaly}

We now turn to present our second approach, which positively characterizes the normal behavior of each application using n-grams as tokens.

A well-known technique for identifying the language of a piece of text is to measure the frequency of the n-grams occurring in the text. An n-gram is a sequence of $n$ (usually consecutive) symbols of the alphabet used in the text (for example, \textsc{ascii} characters, words, etc.). For instance, the most frequent character trigrams in Spanish are \textit{del} and \textit{que}. On the other hand, the character trigrams \textit{whe} and \textit{ike} are very rare in Spanish, but not in an \SQL sentence used for an \SQL-injection attack, as they appear in the keywords \textsc{where} and \textsc{like}. In the most general case, the n-gram may be a sequence of words instead of characters (in which case we may consider the symbols to be the words, and the rest of the analysis remains the same).

\subsubsection{Pre-processing}
The collection of all the n-grams ocurring in a text can be computed following a very simple, fast and linear algorithm, in which a window of size \textit{n} is slided all along the text. This algorithm may be preceeded by a \textit{tokenization procedure}, in which the text is chunked into \textit{words}, separed by a set of \textit{delimiters}. A delimiter is any sequence of symbols matching a given regular expression. In the case of an \URI, for example, the slash symbol (/) and the ampersand symbol (\&) are natural delimiters of an \URI, and the n-grams are sequences of strings, each one describing either the domain of the web site or a resource in its ressource hierachy. For those languages where there are no clear delimiters in the text, words are just the alphabet symbols themselves. Identifying word with symbols provides a tokenization procedure that can be applied in any case, as it is independent from the language grammar. This is an important advantage, as the same method can be uniformelly applied to all fields. Moreover, in practice web applications frequently make use of custom \HTTP fields, and their structure is unknown.


Not all n-grams are equally relevant. Actually, some distinctions may be more problematic than helpful. For example, IP address 168.192.0.1 does not better characterize valid requests than IP 168.192.0.2. On the other hand, making the model to be dependent on the occurrence of particular IP addresses is more prone to overfitting with respect to the training set $\T$. For this reason, we count the n-grams obtained after applying some \textsl{abstraction function} $\textit{abs} : S^m \rightarrow S^k$ on them for some $k\leq m$, where $S$ is the ASCII alphabet. In our experimental results we use an abstraction function performing the following transformations on the decoded \HTTP request: (a) letters are uncapitalized, (b) accents are removed and (c) digits are replaced by the capital letter \enquote{\textit{N}}.

Opposite to the previous approach, in this case the pre-processing step tries to exploit as much as it can from the \HTTP structure. Even if the \HTTP protocol does not impose any structure on the request body, most of the interaction with the user consists in presenting \HTML forms to be filled. 
Once filled, \HTML forms are transmitted to the server as a list of parameter assignments $\text{y}_1=\text{v}_1$ \& $\ldots \text{y}_k=\text{v}_k$ usually separated by the \& symbol. Therefore, the request body may be either plain text or a parameter assignment. In this latter case, we do not work directly on the request string itself as in the one-class approach: we first parse its structure, keeping the \HTTP field and parameter contents, and discarding their names. We use the term \textit{model fields} to design either a header field, or a web application parameter. The model computes a separate signature for each model field.

In the sequel, we introduce a general framework for experimenting with attack detection based on n-gram frequencies. We consider n-grams of length $m$ formed from symbols from the alphabet $S$, for all $m \leq n$, where $n$ is a model parameter that can be adjusted for each application. 

\subsubsection{Feature extraction}

Formally, a model field is a pair formed by an \HTTP request field name and either a parameter name $y_i$ (parameter assignment), or a special value $\bot$, otherwise (for representing unestructured, plain text body). The basic \textsl{attributes} of the model are the n-grams that can be found in the model fields contents. Each attribute is a pair $(x,z) \in \A$ formed with a model field $x$ and an n-gram $z\in S^m$ ocurring in $x$. In the sequel, we use the notation $x.z$ for this pair. The model is enriched with an additional attribute not related to n-grams, namely, the number of characters (length) of each model field $x$. This additional attribute is also a good indicator for code injection attacks, as they are likely to increase the expected field length. 

We associate a random variable $X_a$ to each attribute $a=(x,z) \in A$, which measures events related to the occurrences of the n-gram $z$ in the model field $x$. We focus on two types of events: the number of occurrences of the n-gram in the model field (integer value) or its frequency inside the field contents, that is, the number of occurrences of the n-gram divided into the total amount of n-grams occurring in this model field (real value). Measuring the number of n-grams is better suited for those model fields having an enumerated type of possible values, for example, the languages that the web application support. Measuring the n-gram frequency performs better when the field length may significantly vary from one request to another. For example, a message field in a contact form of the web application contains free text that may be arbitrarily long. The number of occurrences of a given symbol may be quite different depending on the message, but the frequency of that symbol should be rather the same in all messages.

A \textsl{distribution} for one of these random variables $X_a$ is a characterized by a tuple $d = (\mu, \sigma, max, min, N_d, H_d)\in \Distrib$, where $\mu$ is the distribution mean, $\sigma^2$ the distribution variance, $max, min \in \N$ the maximum and minimum values that were sampled for $X_a$, and $N_d\in\N$ the number of sampled values. $H_d : \R \rightarrow \N$ is the histogram counting the number of requests in $\T$ having a given frequency for $a$. 

\subsubsection{Model computation}
We assume that each random variable $X_a$ is independent from the other random variables. This means that constructing the probabilistic distribution of the internal representation of the \HTTP requests amounts to construct one independent distribution for each attribute $a\in\A$. The distribution parameters of each random variable $X_a$ are incrementally approximated using Welford's algorithm for on-line computation of its mean and variance \cite{Welf62}. Therefore, the result of this iterative process is a map $M : \A \rightarrow \Distrib$ representing the learned distribution of the random variable $X_a$ associated to each attribute $a$. This mapping provides the \textsl{language signature} of each model field $x$, in terms of the probability distribution of the n-grams in the specific language of $x$. We compute a specific language for each field. This is an important difference with respect to other works \cite{wang2004anomalous,wang2006anagram}, which analyze the traffic at lower (TCP) layers: while each field represents a logical unit of the web application, with its own specific laguage (numbers, addresses, names, credit card numbers, etc.), this structure is splitted and mixed at the TCP layer into one single language (the union of all these specific languages). 


\subsubsection{Request classification}
In order to test a given \HTTP request $r$, the \WAF computes the map $f_r : \A \rightarrow \R$ of concrete values for each attribute of $r$. Then, each model field $x$ of $r$ is considered. If field $x$ is not defined in model $M$, then $r$ is rejected. This amounts to say that any request containing unknown \HTTP fields or web application parameters that have not been met during the training phase is deemed as anomalous. Otherwise, the \textsl{score} $s_{r.x}=\sum_{i=1..n}\dist_i^x(r,M)$ is computed for field $x$, using a function $\textit{dist}^x_i$.
The definition of the distance depends on the n-gram length. Also, different distances could be associated to different fields. If there is a field $x$ for which the obtained score does not satisfy some given criterion $\C$, the whole request is considered valid, otherwise it is deemed as anomalous (outlier value). The criterion used for determining whether the distance is acceptable is that the score falls inside the rank defined by the minimum and maximum values of the score distribution \textit{ds} associated to the requests in the training set $\T$, that is, $\C(\textit{ds},s_r) = \textit{ds}.min \leq s_r \leq \textit{ds}.max$. This distribution is drawn from the training set $\T$, by computing the score of each individual training request.

The distance from the request signature to the language signature is measured using Mahalanobis distance under the assumption of independence of the random variables associated to the tokens: 
\begin{equation}
\dist^x_i(r,M) = \sqrt{\sum_{z\in \T^x_i} \frac{ | \mu_M(x.z) - f_{r}(x.z) |^2}{\frac{1}{\mid\T^{x}_i\mid}+\sigma^2_M(x.z)}}
\end{equation}
Mahalanobis distance measures how many standard deviations away the request signature is from the language signature. As a consequence, differences for n-grams with highly variable frequencies do not weight as much as differences for n-grams with a rather constant frequency. The constant $\frac{1}{\mid\T_i^x\mid}$ is added to the denominator to prevent a division by zero for those n-grams with constant frequency. Note that the larger the number of samples for $x$'s i-grams is, the greater the difference from the mean weights in the total score.


In practice, the training set $\T$ is a small fragment of all possible \HTTP requests that the web application accepts, so a given n-gram of the test request $r$ may not be present in $\T$, but still be a valid input for the web application. This is problematic, as unknown n-grams are assigned the highest possible anomalyness score. However, most of the fields contain values that are a subset of some larger language. For instance, a form field in the web application corresponding to an address in Madrid is a particular case of a piece of text written in Spanish. Such larger language has its own language signature $P_x$, which can be used as a \textit{prior distribution} for the field $x$. By assigning priors to some fields we decrease the false positive rates: should a given n-gram not be defined in $M$ for that field, we use that prior as the expected distribution for the attribute. Notice that this approach is possible when the analysis is performed at the \HTTP level, when the application fields can be recognized, but is much harder to implement at the lower communication levels. If the n-gram is neither defined in this prior, then we assign the singleton distribution $d$ such that $\mu=\sigma=0=\textit{min}=\textit{max}=0$, $N_d=\mid\T^{x.z}_i\mid$ and $H_d=0\rightarrow \mid\T^{x.z}_i\mid$, which represents that $0$ is the expected frequency for this n-gram according to the training sample $\T$.

In some situations, it may be helpfull to deliberately exclude some fields from the model. For example some web applications generate a dynamic \URI or encode dynamic information as part of the \URI itself. Those \URI have almost no chance of being met during the model training, and therefore are prone to generate false positives. In such cases, excluding the \URI from the analysis improves its performance. Another example that can be excluded are those fields which are supposed to be close to random data, such as a parameter transmitting a passowrd or encrypted data. In our implementation, a configuration file enable to assign the model parameters to be used in the model construction, such as the n-gram length bound $n$,the tokenization method and its delimiter, the distance, the model fields to be excluded from the analisis, etc. Independent values can be assigned for these parameters for each model field.

\section{Experimental results}
\label{section:experiments}
We have evaluated the proposed models experimenting with three datasets. We have privileged the use of public datasets which are intended to be used for analyzing web application attacks.
Unfortunately, there exist very few datasets of \HTTP traffic in general, and even fewer with tagged attacks. The only public datasets that seemed useful for our purpose were the \CSIC dataset \cite{CSIC2010Dataset} and the dataset from the \PKDD challenge \cite{ecml-pkdd-challenge}. We discarded the use of the \DARPA dataset, as it contains \TCP traffic and is no longer representative of the state of the art in nowadays attacks \cite{IngramInoue}. 
We also evaluated the models on a dataset of our own which was obtained from the \HTTP traffic to the web server of a University. We will refer to this dataset as \FING. 

Each of the used datasets is made of a collection of complete \HTTP requests (header and body) which have been partitioned into one training dataset and two testing ones: one for valid traffic and another for anomalous one. In Table \ref{tab:datasetSize}, we present the details of each dataset. As explained before, our models only use normal traffic for training, labeled attacks will be considered only for evaluation purposes. The model performance was measured with the usual true positive rate (TPR) and true negative rate (TNR) indicators. The TPR corresponds to the percentage of anomalous requests that are detected as attacks. The TNR corresponds to the percentage of valid request that are spotted as normal traffic. The results are presented as ROC curves that describe how these two indicators are affected by the parameters of the models. 

The baseline to which we compare the behavior of our models are the TPR and TNR values obtained as the outcome of using \ModSecurity configured with the \CRS version 2.2.9 out of the box.


In the one-class classification approach the main parameter of the classifier is the threshold that governs the size of the clusters. The size is adjusted by a parameter $\lambda$ whose values range from 0 to 1 (see equation \ref{eq:one-class-threshold}). Each value of $\lambda$ determines an operational point of the classifier. For each dataset we shall present:
\begin{inparaenum}[i)] 
\item the ROC curve obtained for the one-class classifier when training is performed using specific data for the web application and varying $\lambda$ in $(0,1]$ (blue curve). This ROC curve can be seen as the ideal result for the classifier since uses an application specific training dataset,
\item the ROC produced with the one-class classifier trained with data from other web applications will be plotted as a green thicker curve. This result addresses the main question in Scenario I. The black diamond over the green curve indicates a default operational point obtained with $\lambda = 0.5$. This point was fixed based on the number of FP for the training set. Since in the Scenario I we do not have specific training data we cannot fine tune the operational point,
\item the result of \ModSecurity using the \CRS out of the box is indicated with a blue square, \item the yellow ball, which shows the performance of the model that combines one-class with \ModSecurity following the combination strategy discussed in Section \ref{improving_modsecurity}
\end{inparaenum}



In the n-gram model approach a fine tuning of the model was experimented, assigning different tokenization methods, distances, prior distributions and n-gram length bounds to each model field. For the sake of simplicity, we restrict the presentation of the results for the n-gram length bounds $n=1\ldots 5$ when an optimized configuration is assigned to the other model parameters.

 
\begin{table}[]
\centering
\caption{Datasets composition}
\label{tab:datasetSize}
\begin{tabular}{l|rrr}
			& Train (Valid) & Test (Valid) & Test (Attack)       \\
	\hline
ECML/PKDD     & 24504        & 10502         & 15110 \\
CSIC          & 36000        & 36000         & 25065 \\
Own Dataset   & 45907        & 19675         & 1287 
\end{tabular}
\end{table} 

\begin{figure*}
    \centering
    \includegraphics[width=\textwidth]{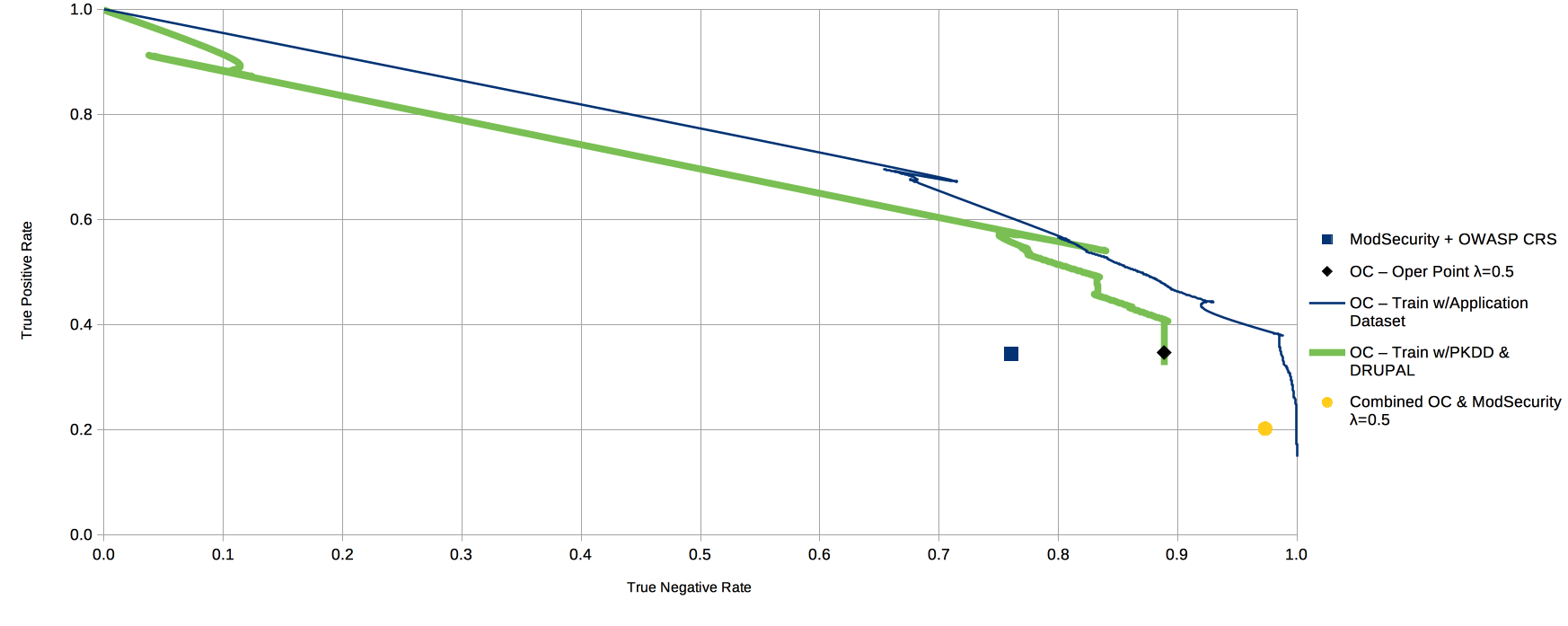}
	\caption{Results for the \CSIC dataset. Blue square: \ModSecurity with \CRS out of the box. Blue solid curve: one-class varying $\lambda$ (OC). Green thicker solid curve: one-class varying $\lambda$ - training with mixed dataset. Black diamond: operational point for OC. Yellow circle: OC operational point combined with \ModSecurity.}
	\label{fig:CSIC2010Dataset}
\end{figure*}
  
\subsection{\CSIC dataset}
In year 2010 the Spanish Research National Council (\textsc{CSIC}) developed and made available a dataset to test a system to protect web applications \cite{CSIC2010Dataset}. This dataset embodies a collection of normal and abnormal \HTTP requests for a web application that provides functionalities to perform on-line shopping on a tiny store. The dataset is made of 36.000 valid requests for the training set, another 36.000 different valid requests for testing normal behavior, and 25.000 abnormal test requests mixing attacks with valid requests containing infrequent characters in the parameter fields (typos). Unfortunately, the distribution between attacks and infrequent values is not specified. The attacks were generated using tools such as Paros (which later became \OWASP \textsc{zed} \cite{Zed}) and w3af \cite{riancho2011w3af}, and include SQL injection, buffer overflow, information gathering, files disclosure, \CRLF injection, \XSS, server side include, parameter tampering, among others. Another important characteristic is that it was conceived and intended for this kind of experiment: the \HTTP requests have a few fields which always take a limited collection of values. It contains several duplicated cases, as each \GET request in the dataset containing a query in the \URI is followed by an equivalent \POST request, with the same query in the body field. The attacks are concentrated on the web application parameters.

In Figure \ref{fig:CSIC2010Dataset} we present the results in terms of TPR and TNR for the \CSIC dataset. The simplicity of the normal requests and the mixture of attack with just anomalous traffic is observed in the high TNR and the low TPR obtained by \ModSecurity (blue square).  As can be noticed, the one-class classifier trained with data from other web applications (green curve), used to evaluate Scenario I, produced good performance scores (TNR and TPR). The ROC curve shows that there are several points that outperform \ModSecurity (blue square). Furthermore, if we compare the results obtained using an application specific training set (blue curve), we can see the performance it is not far from the ideal one (when we train the one-class classifier with data from the same web application). Finally, the yellow ball shows the performance of the model that combines one-class with \ModSecurity. The combination improves in terms of TNR but decreases in TPR. This is because our main objective is to decrease \ModSecurity false positives so the combination algorithm only mark a request to be an attack if both experts tag it as an attack. This means that some requests that were tagged as an attack by the one-class model where tagged as valid by \ModSecurity and viceversa. 

If we compare the results with our baseline, e.g. for the same TNR, \ModSecurity detects around $34\%$ of attacks, where the one-class approach detects around $56\%$. In this particular dataset, the integration of \ModSecurity rules with the one-class approach does not improve the results of one-class by itself in terms of TPR but clearly reduces the number of false positives (i.e., TNR close to 1). See Table \ref{table:ExperimentalResults} for details on the results.

For the n-gram analysis, we configured our tool to perform some specific behaviors for some of the model fields. Despite the default n-gram length, the analysis is always bounded to monogram analysis for \URL, login identifier, customer's national identifier and passwords. This is because the only biased aspect of those fields is the set of allowed characters, but almost any combination of them is possible in principle, so higher order n-gram analysis is prone to produce false positives. In the case of the \URI, the slash bar character (/) is specified as delimiter, and monogram analysis is performed on the number of resulting words that occur in the field, not its frequency. The reason is that is quite unlikely that one of these words (which represent resource folders in the web application directory) appears more than twice in the \URI. Finally, following the technique explained in section \ref{subsection:anomaly}, we specified a prior n-gram distribution for the fields corresponding to names, cities and addresses, drawn from a collection of Wikipedia articles written in Spanish. The use of priors for these fields reduced the false positive rate in 3\% for trigrams. The most significant impact on false positives is observed for bigrams and trigrams. 

\begin{table}[]
\centering
\caption{True negative and true positive rates (in \%) for each dataset}
\label{table:ExperimentalResults}
\begin{tabular}{ll|l|l|l|l|ll}
\cline{3-8}
                                                                    & \textbf{} & \multicolumn{2}{c|}{\textbf{\CSIC}} & \multicolumn{2}{c|}{\textbf{\FING}} & \multicolumn{2}{c|}{\textbf{\PKDD}}                \\ \hline
\multicolumn{2}{|c|}{\textbf{Method}}        & \textbf{TNR}          & \textbf{TPR}         & \textbf{TNR}          & \textbf{TPR}       & \multicolumn{1}{l|}{\textbf{TNR}} & \multicolumn{1}{l|}{\textbf{TPR}} \\ \hline
\multicolumn{2}{|l|}{ModSecurity}                               &            76,1        &         34,3          &         61,1          &           72,2        & \multicolumn{1}{l|}{42,8}  & \multicolumn{1}{l|}{93,0}   \\ \hline
\multicolumn{2}{|l|}{One-class: $\lambda=0,5$}       &            88,9        &          34,6         &            93,3       &                  86,2 & \multicolumn{1}{l|}{0,0}   & \multicolumn{1}{l|}{98,6}   \\ \hline
\multicolumn{2}{|l|}{Combined OC-MS}                     &              97,3      &        20,1           &         99,1          &                  63,0 & \multicolumn{1}{l|}{42,8}   & \multicolumn{1}{l|}{92,1}   \\ \hline
\multicolumn{1}{|l|}{\multirow{5}{*}{N-grams}}          & n=1       &      99,9        &      93,0       &     93,9          &   95,9            & \multicolumn{2}{l}{\multirow{5}{*}{}}             \\ \cline{2-6}
\multicolumn{1}{|l|}{}                                              & n=2       &      99,9        &      94,8       &     94,4          &   97,6            & \multicolumn{2}{l}{}                              \\ \cline{2-6}
\multicolumn{1}{|l|}{}                                              & n=3       &      99,5        &      96,1       &     92,0          &   97,5            & \multicolumn{2}{l}{}                              \\ \cline{2-6}
\multicolumn{1}{|l|}{}                                              & n=4       &      96,2        &      96,8       &     90,7          &   98,8            & \multicolumn{2}{l}{}                              \\ \cline{2-6}
\multicolumn{1}{|l|}{}                                              & n=5       &      90,9        &      97,5       &     89,4           &   98,9            & \multicolumn{2}{l}{}                              \\ \cline{1-6}
\end{tabular}
\end{table}

\begin{figure*}
    \centering
    \includegraphics[width=\textwidth]{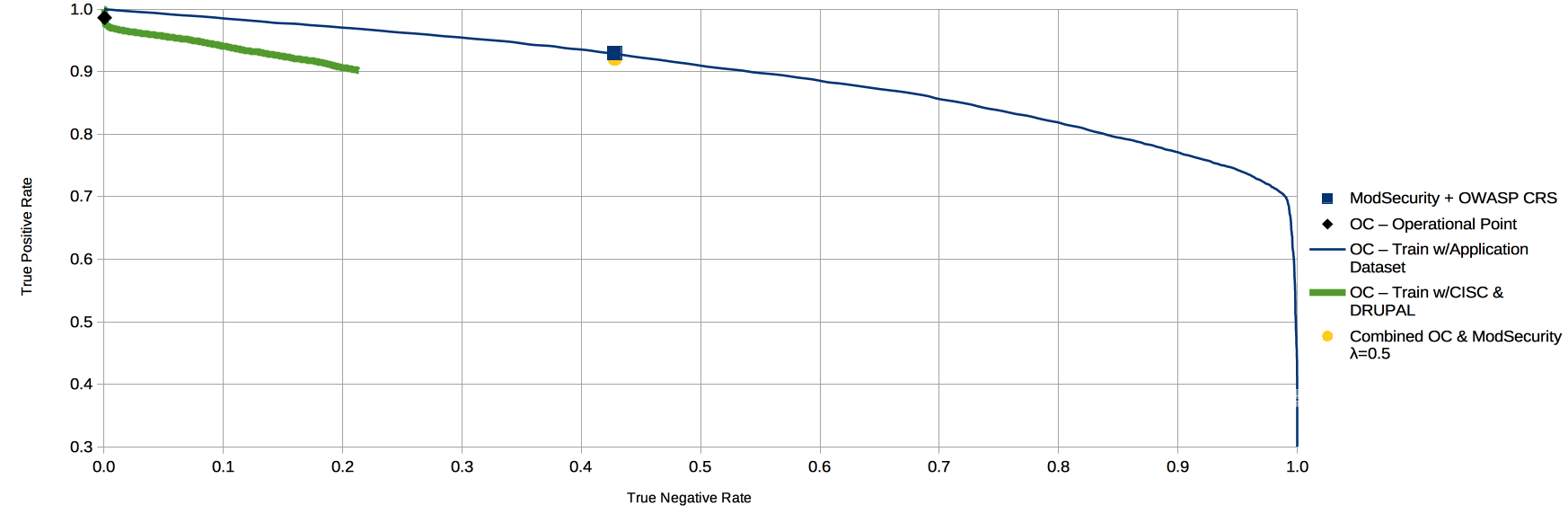}
	\caption{Results for the \PKDD dataset. Blue square: \ModSecurity with \CRS out of the box. Blue solid curve: one-class varying $\lambda$ (OC). Green thicker solid curve: one-class varying $\lambda$ - training with mixed dataset. Black diamond: operational point for OC. Yellow circle: OC operational point combined with \ModSecurity. }
	\label{fig:PKDDDataset}
\end{figure*}

\subsection{\PKDD dataset} 

In year 2007 the 18th European Conference on Machine Learning (ECML) and the 11th European Conference on Principles and Practice of Knowledge Discovery in Databases (PKDD), put forward a challenge on Analyzing Web Traffic \cite{ecml-pkdd-challenge}. The challenge objective was to construct an algorithm based on machine learning techniques to perform multi-class, contextual classification and attack pattern isolation. As part of the challenge it was provided a dataset which contained requests classified in seven different types of attacks plus valid requests. The dataset is composed of 35.006 requests classified as normal and 15.110 requests classified as attacks.

After analyzing the dataset we infer that some process of obfuscation/anonymization was executed on the dataset to protect urls, parameter names and values. Those items were replaced by random values and the obfuscations process does not seem to be an injective function, as no \URI, parameter name or value appears twice in the dataset. As a consequence, the \PKDD dataset revealed useless for carrying on n-gram analysis experiments, as the valid behavior that could be used for training purposes consists in random data with no bias. On the other side, the one-class approach can be successfully used to detect attack signatures. Figure \ref{fig:PKDDDataset} shows the result of this experiment. Regarding the Scenario I (green curve), the performance evaluated with TNR and TPR is not as good as in the case of \CSIC and \FING dataset that will be presented below. After careful evaluation, we observed that when the process of obfuscation is executed in the dataset the random values were generated using letters, numbers and the $=$ and $-$ symbols. These two last symbols are included in our list of tokens defined by the expert. After analyzing the problem, we found that this artificial way of generating the values are so specific and different from the normal traffic to the web application, that the generic approach did not work as expected. On the other hand, if we compare the results obtained using a model trained with specific requests (blue curve), we observe that good performance is achieved. In the curve there are some points with less FP than ModSecurity but slightly worse performance when considering TPR. We believe that the way this dataset was generated strongly conditions the results. Better results should be obtained if we removed the features mentioned above.

\subsection{\FING dataset}
The \CSIC and \PKDD datasets have been artificially conceived for the sake of experimenting with machine learning techniques. In order to evaluate our approach on real life applications, based on actual requests and attacks, we crafted a dataset by registering three days of incoming traffic to the public website of a University \footnote{The dataset is not public, but it is available on demand to other researchers willing to reproduce our results or comparing them against other ones, by writing to the authors}. The only post-processing of this dataset consisted in blurring password values in the request. 

Since the requests are from real traffic, this dataset is less balanced: it has 65582 valid request and 1287 real attacks. It also contains an important amount of custom \HTTP fields used by the applications hosted in the web site. 

The approach of registering traffic in order to exploit it as a dataset has been already used in several previous works. One of the difficulties that raises is classifying the received requests into valid ones and attacks, so that the dataset can be separated into training and a testing part. The web site of the University is protected by an instance of \ModSecurity featuring the \CRS, which has been tuned for several years by a team of security and infrastructure experts. We therefore used \ModSecurity as the labeling tool: those registered requests that were accepted by \ModSecurity were considered as the valid traffic and used for the training step and for the testing dataset, while those requests that \ModSecurity rejected were used as the testing dataset for the attacks. 

\begin{figure*}
    \centering
    \includegraphics[width=\textwidth]{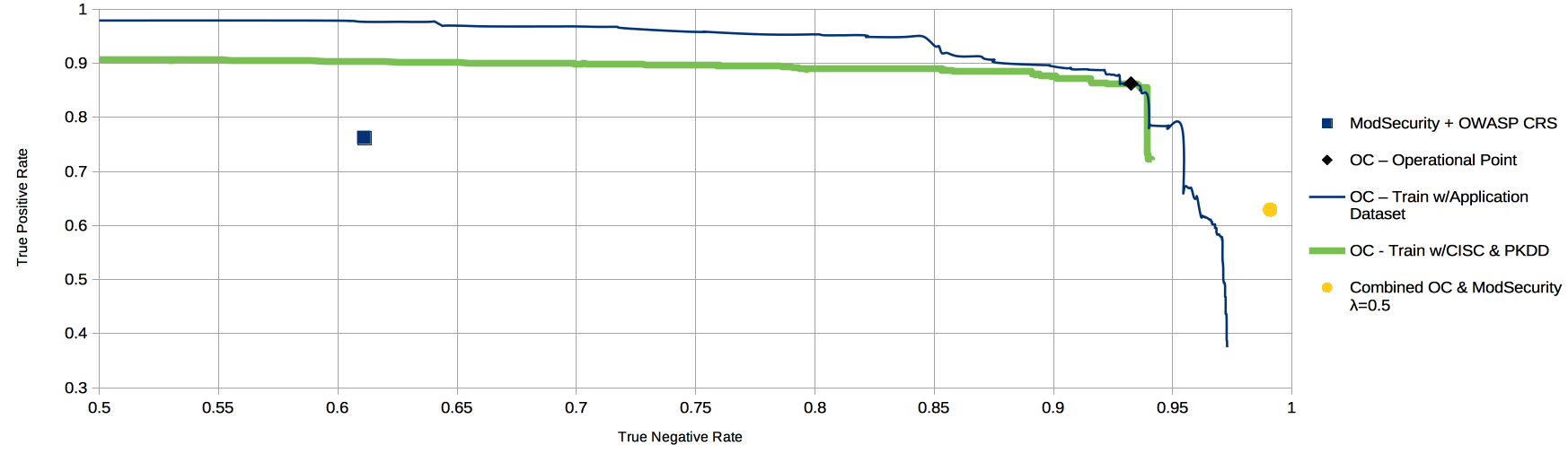}
	\caption{Results for the \FING dataset. Blue square: \ModSecurity with \CRS out of the box. Blue solid curve: one-class varying $\lambda$ (OC). Green thicker solid curve: one-class varying $\lambda$ - training with mixed dataset. Black diamond: operational point for OC. Yellow circle: OC operational point combined with \ModSecurity. }
	\label{fig:FINGDataset}
\end{figure*}

Figure \ref{fig:FINGDataset} describes the results for the \FING dataset. We observe that they are similar to the ones obtained for \CSIC. The one-class classifier trained with data from other web applications (green curve), used to evaluate Scenario I, is very close to the blue curve which is the ROC generated with the one-class classifier trained with application specific data. The default operational point, black diamond, outperforms \ModSecurity, and several of the points in the previous curves clearly perform better than \ModSecurity. Finally, the results of the model that combines one-class and \ModSecurity (yellow ball) improves in terms of TNR but decreases in TPR. This is because our main objective is to decrease \ModSecurity FP so the combination algorithm only marks a request to be an attack if both experts tag it as an attack. This mean that some request that where tagged as an attack by the one-class where tagged as valid by \ModSecurity and viceversa. 



\subsection{Discussion}
Based on the results for the three datasets and the two proposed approaches we discuss in what follows the main questions presented in Section \ref{improving_modsecurity}.

\paragraph{Scenario I: Is it possible to build an attack detection system learning from training data collected from other web applications?} The results for datasets \CSIC and \FING present evidence that it is possible to build a one-class classifier using generic training data, that is, using a dataset with request not from the web application to protect. Furthermore, the degradation with respect to the same classifier trained with specific data for the web application is not critical. For the case of the \PKDD dataset we already discussed the particularities which justify the differences between the one-class classifier trained with specific and non specific training data. 

\paragraph{Scenario I: Can we improve the results of ModSecurity using machine learning methods?} Looking at the results for the datasets \CSIC and \FING we can conclude that it is possible to improve the results of \ModSecurity.  For the \PKDD dataset the results of the method that combines \ModSecurity and one-class are the same as \ModSecurity. One advantage of one-class is that we have several operational points to choose from. In particular, we can reduce the number of FP (increasing TNR) but reducing TPR. As mentioned above, the particularities of the dataset prevent taking advantage of the one-class classifier trained with generic data. For future work we plan to analyze this issue in detail. 

\paragraph{Scenario II: Attainable performance of machine learning methods against ModSecurity} Based on the obtained results we can draw the following conclusions. First, based on the results of one-class and n-grams, it is possible to improve the results of \ModSecurity using machine learning. Second, looking at the TNR and TPR scores computed for the n-gram method, we can say that the n-gram approach is a good solution if we have an application specific dataset to train its model.


\section{Related work}
\label{section:relwork}
There exist in the literature several previous works concerned with the application of machine learning techniques to detect information systems attacks.

\subsection{Virtual patching} 
Web applications are designed using three major tiers: a thin client (web browser), the business logic and a datastore. Virtual patching may be applied to any of those tiers. In recent years work has been reported \cite{BockermannAM09, KarPS161, KarPS162} concerning the use of database firewalls (\DBF) as remediation tools. In all of these works anomaly detection techniques have been applied to analyze the flow of \SQL sentences from the business logic tier of the application to the database with the objective of detecting (only) \SQL injection attacks. As a \DBF can only inspect the traffic between the business logic and the database server it is not capable of detecting attacks that are not directed to the data layer of the application. 

There exists several proprietary implementations of \WAF technology. They have been reviewed and ranked in a recent report of the Garner Magic Quadrant \cite{quadrant2017magic}, where, in particular, one can find the following statement concerning the reviewed technology: \textit{"Use of machine learning is rare and often still unproven"}.


\subsection{Application layer analysis} 
One of the seminal works in anomaly detection techniques was developed by Wang and Stolfo~\cite{wang2004anomalous}, where they introduced \textsc{payl}, a payload-based anomaly detector for intrusion detection. Their approach consists in comparing the byte frequency distribution in network packets using (a simplified) Mahalanobis distance. Accordingly with the current attack vectors of that time, their objective was focused on protecting an internal network from packets carrying worms or other forms of malware. Consistently, their model works on \TCP packets, that is, on the network OSI layer. As the byte distribution heavily depends on the application protocol, they divide the stream into smaller groups of packets, grouping them by destination port and by payload length. They applied PAYL to the traffic to port 80 (\HTTP) of the 1991 \textsc{darpa ids} dataset, obtaining excellent detection results. However, their results relate to the attack vectors of 1999. Indeed, Ingham and Inoue report in \cite{IngramInoue} that the \textsc{darpa ids 99} dataset only contain four web attacks. Moreover, it does not contain any of the \textsc{owasp top 10} attacks \cite{DARPAIDSDataset-attacks}\footnote{This is not surprising, as web applications arose during the 2000, and \SQL injection was first reported in December 1998 \cite{forristal1998}}. As a consequence, while 
\cite{wang2004anomalous} reports on a 100\% detection rate for traffic on port 80 of the \textsc{darpa ids} dataset based on byte frequency distribution of the unparsed \TCP payloads, performing the same analysis on the \textsc{cisc2010} dataset yielded a detection rate of only 40\% of the true anomalous cases. This low performance has been also independently reported by Ingham and Inoue in \cite{IngramInoue}. The reason is that token distribution varies from one \HTTP field or web application parameter to another, which are spread along several \TCP packets. Protecting web applications requires shifting the analysis to the application OSI layer, focusing on \HTTP requests. 

\subsection{Machine learning techniques}

\paragraph{Training from normal traffic only} In the \PKDD challenge \cite{ecml-pkdd-challenge}, the objective was to classify web application requests using a multi-class approach, in which the training dataset contains labeled \HTTP requests which were classified into normal ones and attacks. Several solutions were presented~\cite{exbrayat2007ecml,pkddsummary,gallagher2009classification}. 
In particular, Gallagher et al~\cite{gallagher2009classification} presented a solution that used classical techniques of information retrieval. Even if their solution performs quite well it is restricted to detect only known attacks and relies on a labeled dataset which contains dozens of thousands of requests. In our case, our solutions only require a training dataset that contains normal traffic. In addition to this, we have shown that the one-class approach works quite well  even in the absence of a dataset proper of the application being protected. The scenario in which the \WAF is trained with a dataset containing requests from applications other than the one being protected our results outperform \ModSecurity with the \CRS out of the box. 
In \cite{raissi2007web}, Ra\"issi et al conclude, based on the results of the \PKDD challenge and a feedback survey written by the challengers, that using machine learning techniques to detect web application attacks requires to involve the security experts earlier in the knowledge discovery process. In the one-class approach we have pursued, the feature selection phase uses the knowledge of the security expert to identify the tokens to be considered in the model construction.

\paragraph{One-class classification} This approach models the normal behavior of the application by learning patterns from positive instances. When a new instance arrives, this instance is classified by the model resulting in a score. This score, that is not necessary a probabilistic one, is compared to a threshold to determine whether this new request belongs to the model. This approach, which could also be called \textit{Anomaly}, \textit{Outlier} and \textit{Normality} detection, as far as we know has never been applied to web application protection before.

\paragraph{Analyzing the whole \HTTP request} In \cite{kruegel2003anomaly}, Kruegel and Vigna propose an anomaly detection approach where the attributes of the model capture information about the parameters of the \URI. They focus on characteristics like the parameter length, the order in which they appear and even generate a probabilistic grammar for each parameter. In both of our approaches we analyze the hole request information and not only the \URI parameters. We think this provides further guarantees because all fields are subject to malformed input attacks, as illustrated in the \PKDD dataset \cite{pkddsummary}.  

\paragraph{Token abstraction} Several authors do not directly work on the tokens themselves, but rather on some simplification of them. In \cite{CoronaAG09}, Corona et al. abstract away numbers and alphanumeric sequences, representing each category with a single symbol. Torrano, Perez and Mara\~n\'on \cite{torrano2010anomaly} present an anomaly detection technique where instead of using the tokens themselves, they use a simplification that only considers the frequencies of three sets of symbols: characters, numbers and special symbol, as well as the list of special symbols used in each field. In general, what is counted is the number of symbols after applying some forget functor on the set of tokens, which abstracts away irrelevant differences. Removing information simplifies and speeds up the analysis by reducing the combinatory of possible tokens. It also reduces the risk of overfitting. However, if too much information is removed from the tokens some attacks may become indistinguishable from correct behavior, so an adequate compromise shall be found. In our experiments, the forget functor performs the following transformations: (a) letters are transformed into lowercase and (b) accents are removed (for instance, the letter \textit{\'a} is transformed into \textit{a}). In addition to this, in the n-gram approach, digits are replaced by the capital letter \enquote{\textit{N}}. Such transformations provide a good balance, colapsing all the possible variants of the \textsl{select} keyword using in \SQL injection attacks (namely, Select, SeLeCt, etc.) but still enabling to differentiate this keyword from other pieces of natural language text which may be part of text field in a normal request.

\paragraph{N-grams made of words} The contents of each request field is a string, so most of the anomaly detection approaches make use of document classification, \NLP and information retrieval techniques. As content languages may significantly vary from one application to another (and even from one \HTTP field to another), in most of the previous work tokens are just n-grams 
\cite{wang2004anomalous,wang2006anagram,kruegel2003anomaly,torrano2010anomaly,Corona08,CoronaAG09,closeLookToNgrams}. An n-gram is usually a sequence of consecutive characters, even though \cite{Corona08} shows that using $n$ characters which are $m$ positions away one from the other may also be an alternative. We generalize and extend the idea of n-grams and words in the following way: the contents string is first split into a sequence of sub-strings (tokens) separated by other sub-strings meeting a regular expression (the \textsl{delimiters}). Then, we consider n-grams formed with sequences of these tokens. This enables a more accurate analysis of some \HTTP fields such as the \URI, where there exists a clear delimiter character (the slash) which enables to divide the contents into tokens, namely, the words of the directory resources. Other fields, such as \texttt{Accept-Encoding} or \texttt{Accept-Language} also make use of commas and semicolons as delimiters, and can be addressed in terms of words instead of characters. 

\paragraph{Fine granularity in the analysis} Among the works focusing on the application layer, a few of them consider each \HTTP request as a single unstructured document to be analyzed and classified \cite{gallagher2009classification,ecml-pkdd-challenge}. This approach is prevalent when focusing on the current state of the art in attacks, and is also the approach proposed in the one-class classification technique presented in this paper. Other works perform some partial parsing of the \HTTP structure, for example in \cite{EstevezTapiador2004175,kruegel2003anomaly}. However, in this latter works, the parsing is reduced to the structure of the \URI, splitting the \URI into the URL and the web application parameters. In our proposal for n-gram analysis we go further, splitting the request contents into the contents of its header, header fields, and web application parameters (either appended to the \URI or in the \HTTP body). We advocate for this approach because each field has its own independent language, which may be quite different from the language of other fields. As a consequence, a much more specific and biased contents can be extracted from each one than from the whole request.

\paragraph{Higher order n-grams} Initial attempts for using n-gram analysis on \HTTP requests focused on monograms and single character distributions \cite{wang2004anomalous,torrano2010anomaly}. However, Kolesnikov and Lee \cite{Kolesnikov04advancedpolymorphic} and  Pastrana et al. \cite{Pastrana2015} have shown that short n-grams are vulnerable to mimicry attacks, in which the attacker carefully adds characters in order to get closer to the n-gram distribution expected by the \WAF. Mimicry attacks become much more challenging when using higher n-grams \cite{wang2006anagram}. Opposite to the cited proposals, we do not use n-grams of a fixed length $n$, but all the n-grams of any length up to a given model parameter $n$ that can be adjusted for each case. Moreover, in our proposal the n-gram length may be configured for each field. In this way it is possible to set $n=3$ for all the model fields, but restrict the analysis to monograms on cookies, email addresses or other fields where higher order n-grams are too sparse but not all symbols are allowed.



\section{Conclusion and Further Work}
\label{section:conclusion}
%
%
%

We have presented two complementary approaches to enhance the detection capabilities of the \ModSecurity \WAF. 

We follow a one-class classification approach to learn the behavior of normal web application requests based on features that represent the payload of an attack. Requests with abnormal behavior are classified as attacks. This classification process draws its model attributes from the knowledge that the security experts have concerning the current state of the art in web application attacks. The values of those attributes are then trained for nominal cases and  learned from the normal behavior of the application. This classification approach provides the means for addressing attacks performed on fields which are expected to contain weakly biased data. 

We also put forward an anomaly detection approach based on the expected n-gram frequencies for the fields and web parameters of the application. For that we make use of higher order n-grams of multiple lengths. We have experimented with several variants of this general framework and described results obtained with some particular instances of the model parameters. Anomaly detection draw its attributes from the expected normal behavior of the application, and deems as an attack those \HTTP requests which deviates from such behavior. This positive characterization of the normality makes the \WAF more resilient to zero-day attacks. 


Experiments have been performed on three datasets of very different nature. The results on these datasets show the potential of machine learning approaches in the development of web application firewalls. The proposed method using a one-class classifier provide better detection and false positive rates than \ModSecurity configured with the \CRS out of the box. Regarding the n-gram approach, we show that in the two datasets used for testing it, the results clearly outperform the ones of \ModSecurity. As we said before, the two approaches can be seen as complementary, the one-class can be used in Scenario I, when no specific training data is available, and the n-gram is suitable for Scenario II since it needs specific training data to learn from normal traffic.


A mid-term goal is to implement a module enabling to introduce machine learning techniques into ModSecurity itself. In the meantime, the prototypes implemented for these experiments can be immediately used to shorten the time required for tuning the Core Rule Set for a particular web application: running them on ModSecurity's logs and analyzing those \HTTP requests that are deemed normal by the tools to spots false positive examples that can be used as basis to tune the rules. 

Regarding the combination of one-class and \ModSecurity, in this work we proposed a simple combination based on the output labels of both components (normal, attack). In future work we will study the combination of the scores provided by both of them. These scores contain more information than just the classification outcome. In the area of classifier combination, there are several alternatives to improve the classification fusing the scores, or posterior probabilities, of multiple classification systems.

An alternative approach to address random contents fields is to apply a symmetry approach to anomaly detection, namely, use n-grams to learn the language signature of the malicious payloads for each attack technique in the current state of the art, and deem as normal the contents of any field that do not match such signature.


In order to push further the use of machine learning techniques in \WAF technology, there is a strong need for disposing of real, open, free, standardized and documented datasets of several web application behavior. Such datasets shall offer a documented description of what web application functionalities are covered in the dataset, and the type of attack payloads that were used for testing. They shall also include erroneous or atypical input which does not correspond to attacks, in order to test the actual TNR that the ML technique offers.  Should an anonymization procedure be required to publish the dataset, it shall respect the structure and correlation of original data. These goals can only be achieved by automated tools for building such datasets for a given application. Providing such tools is part of our further research agenda.
%
%
\bibliographystyle{IEEEtran} 
\bibliography{wafintl}
%
\end{document}